\documentclass[aps,prl,twocolumn,showpacs,preprintnumbers,amsmath,amssymb,superscriptaddress]{revtex4}
\usepackage{graphicx}
\usepackage{dcolumn}
\usepackage{bm}

\begin{document}

\title{Isolated Vibrational Wavepackets in D$_2^+$:
Defining Superposition Conditions and Wavepacket
Distinguishability}
\author{W. A. Bryan}
\altaffiliation[Also at ]{Central Laser Facility, Rutherford
Appleton Laboratory, Chilton, Didcot, Oxon OX11 0QX, UK}
\email{w.bryan@ucl.ac.uk} \affiliation{Department of Physics and
Astronomy, University College London, Gower Street, London WC1E
6BT, UK}
\author{J. McKenna}
\affiliation{Department of Pure and Applied Physics, Queen's
University Belfast, Belfast BT7 1NN, UK}
\author{E. M. L. English}
\affiliation{Department of Physics and Astronomy, University
College London, Gower Street, London WC1E 6BT, UK}
\author{J. Wood}
\affiliation{Department of Physics and Astronomy, University
College London, Gower Street, London WC1E 6BT, UK}
\author{C. R. Calvert}
\affiliation{Department of Pure and Applied Physics, Queen's
University Belfast, Belfast BT7 1NN, UK}
\author{R. Torres}
\affiliation{Blackett Laboratory, Imperial College London, Prince
Consort Road, London SW7 2BW, UK}
\author{D. S. Murphy}
\affiliation{Department of Pure and Applied Physics, Queen's
University Belfast, Belfast BT7 1NN, UK}
\author{I. C. E. Turcu}%
\affiliation{Central Laser Facility, Rutherford Appleton
Laboratory, Chilton, Didcot, Oxon OX11 0QX, UK}
\author{J. L. Collier}
\affiliation{Central Laser Facility, Rutherford Appleton
Laboratory, Chilton, Didcot, Oxon OX11 0QX, UK}
\author{J. F. McCann}%
\affiliation{Department of Pure and Applied Physics, Queen's
University Belfast, Belfast BT7 1NN, UK}
\author{I. D. Williams}%
\email{i.williams@qub.ac.uk}%
\affiliation{Department of Pure and Applied Physics, Queen's
University Belfast, Belfast BT7 1NN, UK}
\author{W. R. Newell}%
\email{w.r.newell@ucl.ac.uk}%
\affiliation{Department of Physics and
Astronomy, University College London, Gower Street, London WC1E
6BT, UK}
\date{\today}

\begin{abstract}
Tunnel ionization of room-temperature D$_2$ in an ultrashort (12
femtosecond) near infra-red (800 nm) pump laser pulse excites a
vibrational wavepacket in the D$_2^+$ ions; a rotational
wavepacket is also excited in residual D$_2$ molecules. Both
wavepacket types are collapsed a variable time later by an
ultrashort probe pulse. We isolate the vibrational wavepacket and
quantify its evolution dynamics through theoretical comparison.
Requirements for quantum computation (initial coherence and
quantum state retrieval) are studied using this well-defined
(small number of initial states at room temperature, initial
wavepacket spatially localized) single-electron molecular
prototype by temporally stretching the pump and probe pulses.
\end{abstract}

\maketitle

As with a broad spectrum of quantum systems, if a molecule
laser-induced into a coherent superposition of vibrational states
is free to evolve, a time-dependent wavepacket is created. The
wavepacket spreads or dephases significantly after the initial
superposition, the result of the constraining potential surface
anharmonicity. However, when the discrete eigenstates naturally
rephase or revive, the original state of the system is almost
fully recovered, and the initial quasi-classical periodicity
returns. This phenomenon of wavepacket revival \cite{Rob04} has
major implications in experimentally realisable time-evolving
quantum systems. Through the precise optical preparation of a
tailored wavepacket in a molecule, the wavepacket motion is
imprinted on the dissociative states, observable with a
probability governed by the amplitudes that contribute to the
superposition.

In this Letter, we report the isolation and quantification of a
vibrational wavepacket in a room-temperature ensemble of D$_2^+$
ions. We vary the initial coherence of the superposed states and
distinguish structure in the wavepacket revival when the
measurement condition is changed. An application of these results
is as a single-internal degree-of-freedom {\it prototype}
vibrational qubit (vibrational states, $\it v$, support the
qubit), the building block of a molecular quantum computer (MQC).
Quantum computation operations are not performed, rather we test
the preparation and observation of a superposition. While MQC has
been demonstrated in Li$_2$ \cite{Vala02}, the D$_2^+$ system is
considered as it is readily handled theoretically (numerical
solution of the rotation-free 1D TDSE under the two-state
approximation) \cite{Dom07} and generated experimentally (D$_2$
molecules at room temperature are predominantly in {\it v} = 0, no
state-selective cooling is required; short laser pulse defines a
spatially localized initial state) \cite{Dom07,McK06,Erg06}.

Before implementing quantum computation (QC), five general
`physical implementation' requirements have been identified
\cite{DiVin00}: (1) the ability to initialize the qubit in a
fiducial state, (2) the coherence time of the qubit is much longer
than the operation times, (3) a universal set of logic operators
are available, (4) a qubit-specific measurement capability exists
and (5) the quantum system is well characterized and readily
scalable. We quantify the fulfilment of requirements (1), (4) and
(5) with a discussion of requirement (2).

Clearly requirement (3) lies at the heart of QC, which in future
experiments will be attempted by optically controlling the phase
\cite{Lee04} or amplitude \cite{Niik04} of each vibrational state,
referred to as phase or amplitude manipulation. The additional
degrees of freedom in a larger molecule could be employed
\cite{Viv02}, i.e. compound vibrational or bending modes, making
scalable (multi-dimensional) QC feasible.

\begin{figure}
\includegraphics[width=240pt]{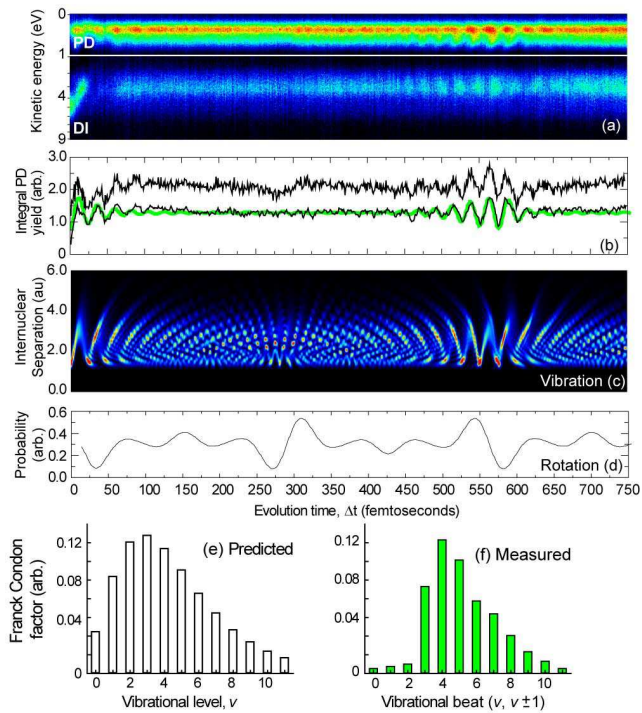}\\
\caption{(Color online) Comparison between experiment and theory
showing vibrational and rotational revivals. (a) D$^+$ PD (top)
and DI (bottom) yield as a function of wavepacket evolution time,
$\Delta$t, and ion kinetic energy, recorded at room temperature.
(b) Integrated PD yield (thick black) contains vibrational and
rotational revivals (top). The theoretical prediction (green /
thick grey curve) shows remarkable agreement with the experimental
results when rotational dynamics are excluded (thin black curve).
(c) The calculated time-dependent probability distribution. (d)
The calculated impulsive alignment of D$_2$ induced by the pump
pulse in a room-temperature distribution of rotational levels. (e)
The predicted Franck-Condon (FC) distribution. (f) Fourier
Transform of the experimental PD yield. The frequency components
present in the vibrational wavepacket are beats between
vibrational states $\it{v}$ and $\it{v}\pm 1$. The Fourier
amplitudes thus return a measure of vibrational
population.}\label{fig1}
\end{figure}

Tunnel ionization (D$_{2}$ ($\it{v'}$ = 0) + {\it pump}
$\rightarrow$ D$_2^+$ ($\it{v}$ = 0, 1, 2...) + e$^{-}$) when the
Keldysh parameter $\gamma$ $<$ 1 \cite{Pop04} creates a
vibrational wavepacket on the 1s$\sigma_g$ potential surface of
D$_2^+$ \cite{Dom07,McK06,Erg06}, and is near-instantaneous so is
treated as a `vertical' transition. The configuration of the
potential energy surfaces is such that ionization only transfers
population around the inner turning point of the wavepacket motion
\cite{Dom07,McK06,Erg06}. In a phase manipulation MQC, such a
transition fulfils requirement (1), as all $\it v$-states are
initiated with essentially zero phase difference. In an amplitude
manipulation MQC, requirement (1) is similarly fulfilled as all
accessible $\it v$-states have a known (Franck-Condon) amplitude
distribution \cite{Urb04}. However, in practice an ensemble of
molecules may be used to generate meaningful signal, where it
becomes important that the pulse length is sufficiently short to
retain coherence in the ensemble.

Observing the vibrational wavepacket evolution on the 1s$\sigma_g$
ground potential surface is possible by `collapsing' the
wavepacket to an observable state with a second laser pulse.
Photodissociation (PD) D$_2^+$ + {\it probe} $\rightarrow$ D$^+$ +
D breaks the internuclear bond via photon coupling the
1s$\sigma_g$ ground and 2p$\sigma_u$ first excited states
\cite{Post04}. A second route to observing wavepacket evolution is
through double ionization (DI) D$_2^+$ $\rightarrow$ D$^+$ + D$^+$
+ e$^-$, requiring a higher laser intensity as a second ionization
event must occur \cite{Post04}. PD and DI are distinguishable by
the kinetic energy released during bond fracture. Both PD and DI
are enhanced when the wavepacket is at the outer turning point
$\simeq$ 3.5 a.u., \cite{Dom07,McK06,Erg06}, and the internuclear
potential energy surfaces naturally act as a quantum `shutter' or
`barrier' depending on whether they are laser-coupled or not.

The experimental set-up is an extension of one described elsewhere
\cite{Dom07,McK06,Bry07}: intense (4 $\times$ 10$^{14}$
Wcm$^{-2}$) ultrashort (12 femtosecond, 1 fs = 10$^{-15}$ s) near
infra-red (800 nm) laser pulses are used to create (pump) and
observe (probe) the vibrational wavepacket. A co-linear
low-dispersion Mach-Zehnder interferometer capable of supporting
the pulse bandwidth ($>$ 100 nm) generates the time-separated pump
and probe, allowing a delay 0 $\leq$ $\Delta$t $\leq$ 160
picoseconds between the pump and probe with a repeatable
resolution of $\simeq$ 300 attosecond (1 as = 10$^{-18}$ s). An
alternative method for generating two or three time-delayed pulses
has been demonstrated \cite{Lee06} by transmitting a $\simeq$ 10
fs pulse through a disc-double annulus optic allowing two variable
delays of hundreds of fs if the two annuli are radially rotated.
After focussing with a spherical mirror, the interaction volumes
generated by our Mach-Zehnder configuration overlap perfectly,
whereas in the disc-annulus configuration, a time-varying spatial
overlap is produced. Furthermore, the dispersion of the
Mach-Zehnder configuration does not change with delay. D$^+$
fragments from PD and DI, Fig. 1(a), are detected in an ion
time-of-flight mass spectrometer containing a 250$\mu$m aperture,
thus signal can only result from PD or DI when the molecular axis
is parallel to the probe polarization and detector axis. The
aperture also defines a quasi-isointensity source in the focal
volume.

In addition to creating a D$_2^+$ vibrational wavepacket,
few-cycle pulses impulsively align deuterium \cite{Bry07,Lee06},
thus the revival of a rotational wavepacket is expected. We
isolate the vibrational and rotational motions, which is only
possible for a limited range of pump pulse durations as discussed
later. As shown in Fig. 1(a), in the region 450 $\leq$ $\Delta$t
$\leq$ 650 fs, both PD and DI yields exhibit a $\simeq$ 24 fs
revival modulation with maximum contrast at 550 fs, the result of
vibrational revival, consistent with recent findings
\cite{Dom07,McK06,Erg06}. However, between 250 $\leq$ $\Delta$t
$\leq$ 350 fs a rotational revival is also observed, comparable to
recent observations \cite{Bry07,Lee06} which will also underlie
the vibrational revival around 550 fs. Both the rotational and
vibrational revivals are evident in the integrated PD yield as a
function of $\Delta$t, Fig. 1(b) (top).

A major finding of the present work concerns the separation of
these two wavepackets. The isolation of the vibrational wavepacket
requires confirmation that two distinct populations in D$_2$ and
D$_2^+$ exist following the pump. In recent work by the authors
\cite{Bry07} (see also \cite{Lee06}), we quantify the rotational
wavepacket in D$_2$ by Fourier transformation of the high-energy
time-dependent DI signal recorded under identical experimental
conditions. We observe peaks (with an accuracy of 5\%) in the
frequency spectrum at 5.5 (0), 9.2 (1), 12.8 (2), 16.4 (3), 20.1
(4) and 23.7 (5) THz, where parenthesis indicate the J : J + 2
rotational state beat in the D$_2$ molecule. There is no evidence
for a rotational wavepacket in D$_2^+$, which would contribute
peaks in the frequency spectrum at 2.7 (0), 4.5 (1), 6.3 (2), 8.1
(3), 9.9 (4), 11.7 (5), 13.5 (6), 15.3 (7) and 17.1 (8) THz. This
is confirmed in the present work. Furthermore, the vibrational
wavepacket can only exist in the D$_2^+$ molecular ion generated
in the pump, as the pump duration is greater than the
quasi-classical period (11 fs) of D$_2$ as recently investigated
\cite{Ergl06}.

Thus we find that the pump either creates a vibrational wavepacket
in D$_2^+$ (rotating in a non-time-correlated manner with respect
to the pump with a room-temperature distribution of J-states) or a
rotational wavepacket in D$_2$ (vibrating incoherently in an
impulsively modified distribution of J-states). Solving the TDSE
for the vibrational \cite{Dom07} (Fig. 1(c)), and rotational
\cite{Bry07} (Fig. 1(d)) degrees of freedom, we quantify these
motions. We take the integrated PD yield as a function of
$\Delta$t (Fig. 1(b) top), and subtract the probability of finding
the rotational wavepacket within the detector acceptance, Fig
1(d). The integrated PD yield (arb. units) is scaled relative to
the rotational wavepacket probability until the signature of the
rotational wavepacket (250 $\leq$ $\Delta$t $\leq$ 350 fs) is
totally suppressed. The resulting PD yield contains the pure
vibrational wavepacket, as shown in Fig. 1(b) (bottom). The PD
yield is modelled by numerically propagating the initially
localized state (FC principle) then calculating the time dependent
coupling of the 1s$\sigma_g$ and 2p$\sigma_u$ states \cite{Dom07}.
The vibrational wavepacket in the molecular ion is clearly
well-described by this relatively simple treatment, thus
requirement (5) is partially fulfilled. If evidence for strong
rotational-vibrational coupling had been found, this would not be
the case.

\begin{figure}
\includegraphics[width=230pt]{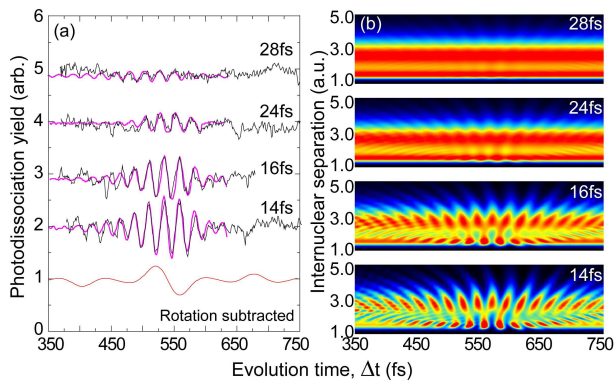}\\
\caption{(Color online) Controlling the initial superposition in
an isolated vibrational wavepacket. (a) Experimental PD yield as
the pump duration is increased (thin black curve). The rotational
wavepacket has been removed from the dataset as discussed. The
shape of the revival indicates the superposition degradation as
the pump is stretched, illustrated by comparison with a
theoretical prediction (thick magenta / grey curves). (b) The
predicted vibrational wavepacket after averaging over the changing
Gaussian pump duration.}\label{fig2}
\end{figure}

Fourier transformation of the raw time-dependant PD yield, Fig
1(b), recovers the beat frequencies of the vibrational levels with
the associated amplitudes that form the vibrational wavepacket,
therefore we have a method for recovering the population of the
vibrational states. As is apparent from Fig. 1(e) and (f), the
measured distribution of vibrational beat frequencies agrees with
the expected FC distribution of states for $\it v$ $>$ 4. For
states $\it v$ $<$ 4, negligible beating is observed. This is a
result of the probe pulse being of insufficient intensity to open
the three-photon crossing \cite{Post04}, thus the vibrational
wavepacket `below' v = 4 remains bound. Our `quantum shutter' is
therefore selective, however the principle is sound, since given a
probe intensity $>$ 6 $\times$ 10$^{14}$ Wcm$^{-2}$ all {\it v}
states will be accessed. It should be noted that the states 0 $<$
$\it v$ $<$ 4 must be populated, otherwise the quasi-classical
period of D$_2^+$ would not be observed: lack of population in
these states would increase the vibrational revival periodicity.
In a phase manipulation MQC, the relative phase can be temporally
resolved for occupied {\it v}-states; while in an amplitude
manipulation MQC, the relative amplitudes of the constituent {\it
v}-states can also be recovered by Fourier transform analysis,
Fig. 1(f). Returning to QC requirement (4), the ability to
determine the final state in our measurement is met. Destruction
of the prototype qubits is necessary to read the final state,
however it is not necessary to destroy the wavepacket to
manipulate. By utilizing nondestructive pulses, it should be
possible to modify the wavepacket while maintaining the
superposition.

To extend our understanding of fulfilling requirements (1) and
(4), we quantify variations in the initial vibrational wavepacket
coherence. We also study wavepacket distinguishability, the
faculty to temporally discern subsequent oscillations of the
vibrational wavepacket, particularly around revival. Both are
macroscopically achieved through independently varying the pump
and probe durations by changing the dispersion (hence chirp) in
either arm of the interferometer.

Fig. 2(a) shows the integrated PD yield in the vicinity of the
vibrational revival as the pump duration, $\tau_{pu}$, is
increased from 14 to 28 fs while the probe duration, $\tau_{pr}$,
is fixed at 13 fs. The vibrational wavepacket in D$_2^+$ was
isolated from the rotational signature in D$_2$ as described
earlier. The contrast of the vibrational revival structure
decreases as the pump duration increases, the consequence of loss
of coherence in the initial superposed state. The experimental
revival structure (thin black curve, Fig. 2(a)) is compared to the
theoretical wavepacket (Fig. 1(c)) averaged over the pump pulse,
Fig. 2(b), and integrated 3 $\leq$ R $\leq$ 10 a.u., then overlaid
(thick magenta / grey curves) on the experimental data in Fig.
2(a). The resultant magenta curves show an excellent agreement
with the decohering isolated vibrational wavepacket, clearly
demonstrating the role of coherence in the initial pump pulse.

In Fig. 2(a), it is apparent that even for $\tau_{pu}$ = 28 fs,
some vibrational coherence remains when $\tau_{pu}$ is longer than
the revival oscillatory period, 24fs. This residual structure is
due to the highest vibrational states, {\it v} $>$ 9 which have
the longest periods and an absolute population of less than 4\%.
Consequently, care must be taken in defining the initial state of
a vibrational qubit as states with an apparently negligible
population clearly influence wavepacket evolution.

By increasing $\tau_{pr}$ from 14 to 60 fs while keeping the pump
fixed at $\tau_{pu}$ = 14 fs (thus the vibrational wavepacket is
identical to Fig. 2(b), bottom, throughout), the amplitude of the
revival changes suddenly as $\tau_{pr}$ exceeds 40 fs. Since a 15
fs probe, Fig. 3(b), is short compared to the revival,
oscillations are observed with high fidelity. At 30 fs, Fig. 3(c),
a comparable fidelity is also achieved, as subsequent revivals are
still distinguished. However, a 45 fs probe, Fig. 3(d), cannot
distinguish any two revival maxima, hence the oscillatory
structure is suddenly lost when $\tau_{pr}$ approaches double the
quasi-classical period of D$_2^+$. We can therefore define a
wavepacket distingushability condition, which is less strict than
for creation, depending on the measurement regime. If a
time-dependent measurement is performed at revival (as required in
the phase manipulation), $\tau_{pr}$ must be shorter than double
the quasi-classical period to ensure total wavepacket collapse.
Alternatively, to guarantee that all the wavepacket is measured in
an energy-resolved scheme (as required in the amplitude
manipulation), $\tau_{pr}$ of at least double the period of the
highest populated state is required. Furthermore, the intensity
requirements discussed earlier must also be adhered to.

\begin{figure}
\includegraphics[width=230pt]{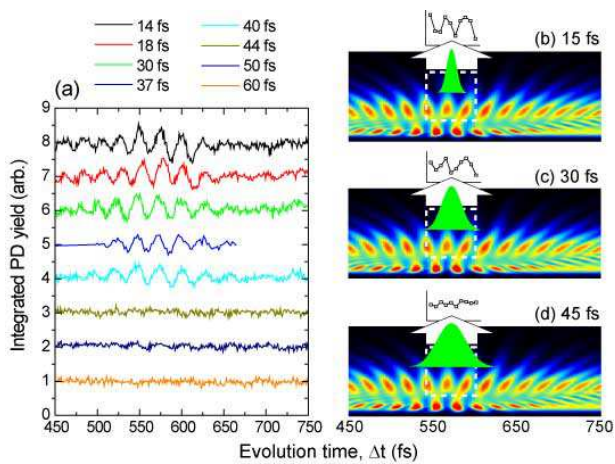}\\
\caption{(Color online) Distinguishing the reviving vibrational
wavepacket as the measurement condition is changed. (a) The sudden
loss of revival fidelity when the probe duration exceeds 40 fs,
the consequence of the temporal structure becoming
indistinguishable to the probe. From top to bottom, the probe
durations are indicated sequentially. Comparing a (b) 15 fs, (c)
30 fs and (d) 45 fs probe to the revival structure, the facility
to decern subsequent oscillations (i.e. wavepacket
distinguishability) of the D$_2^+$ vibrational revival is
compromised.}\label{fig3}
\end{figure}

The pump-variation observations indicate a promising future for
controlling the initial superposition conditions. Following the
optical preparation of an isolated vibrational wavepacket in a
single molecule, a genetic algorithm will determine the
intermediate time- and frequency-structured laser pulse
\cite{Lev01} required to achieve a particular target state (either
by phase or amplitude manipulation). This intermediate `operator'
pulse must be low intensity so as not to fracture the internuclear
bonds, rather to guide wavepacket evolution. The high intensity
ultrashort probe pulse will then fracture the internuclear bonds,
allowing the result of the operation to be read.

To perform logic operations and apply successful error correction
schemes requires a coherence time $\simeq$ 10$^4$ times the
interaction time, t$_I$, requirement (2). Here, ionization defines
t$_I$ $\simeq$ 1.3 fs thus if the wavepacket survives to $\simeq$
10 ps with high fidelity, QC is possible. Our measurements have
exceeded 2 ps, therefore this constraint does not appear
technologically limiting. The MQC will be scaled by increasing the
number of vibrational modes supporting the qubit, accomplished in
a larger molecule. Communication between states will then be
controlled optically by distorting the electronic potential
surface through the Stark-shift.

In a broader application, a pump of a shorter duration than the
vibrational motion establishes a time-dependent variation of the
refractive index in a Raman-active molecular medium, as
demonstrated in SF$_6$ \cite{Zhav02}. By employing the vibrational
mode of a hydrogenic molecular ion, it may be possible to
spectrally broaden and self-compress, potentially to the
single-cycle level.

The experiment was carried out at the ASTRA Laser Facility, STFC
RAL, UK. The assistance of J. M. Smith, E. J. Divall, K. Ertel, O.
Chekhlov, C. J. Hooker and S. Hawkes is gratefully acknowledged.
This work was funded by the Engineering and Physical Sciences
Research Council (UK). RT acknowledges the Spanish Department of
State of Education and Universities, and the European Social Fund.

\end{document}